\documentclass[aps,pra,amssymb,twocolumn, amsfonts,amsmath,showpacs, superscriptaddress]{revtex4}
 \usepackage{graphicx}
  \usepackage{amssymb}
  \begin{document}
\title{Chronogenesis, Cosmogenesis and Collapse}
\author{Philip Pearle}
\affiliation{Emeritus, Department of Physics, Hamilton College, Clinton, NY  13323}
\begin{abstract}
 {A simple quantum model describing the onset of time is presented. This is combined with a  simple quantum model of the onset of space. A major purpose  is to explore the interpretational issues which arise.  The state vector is a superposition of states representing ``instants."   The sample space and probability  measure are discussed. Critical to the dynamics is state vector collapse:    it is argued that a tenable interpretation is not possible without it.   Collapse provides a mechanism whereby  the universe size, like a clock,  is narrowly correlated with the quantized time eigenvalues.}  
 \end{abstract}
 \email{ppearle@hamilton.edu}
\maketitle

\section{Introduction}\label{sec1} 

The purpose of this paper is to provide a simple quantum model describing the onset of time and space from an initial ``nothingness" state $|0\rangle$, and to investigate issues arising from it. 

 Inflationary models do not start from such a state.  They take as starting point an incredibly small universe which, nonetheless, has evolved  enough so that it can be described by a quantum field  on a classical general relativistic background\cite{inflation}.  
 
In the model given here,  since there is no ab initio variable $t$,  there is no Schr\"odinger equation.  Instead,  the state vector $|\Psi\rangle$ is given by what might be called an existential equation, 
\begin{equation}\label{1}
|\Psi\rangle=V|0\rangle.
\end{equation}
\noindent  $V$ is a (non-unitary) operator to be specified (section \ref{sec2}) which depends upon a time operator $T$.  This differs  from the Wheeler-DeWitt equation whose form is $H |\Psi\rangle =0$ and which has no time operator, so time has to be somehow constructed\cite{Wheeler-DeWitt}.  This model does not have the baggage of General Relativity, although that may be added.

Conceptual alternatives are that time does not exist\cite{Barbour} or that time has been going on forever\cite{Penrose}.  However, time may exist and may not have been going on forever\cite{timeis}. 

Since equations describing dynamical wave function collapse shall be used, it is necessary to give a brief  overview of why standard quantum theory needs to incorporate collapse and the means of  accomplishing this.  

\subsection{Non-Collapse Quantum Theory}\label{sec1A}

In the standard quantum description,  existence of classical time $t$ is assumed.  The state vector $|\psi,t\rangle$ evolves with time from an initial state $|\psi,0\rangle$ according to the unitary Schr\"odinger  evolution  $|\psi,t\rangle=U(t)|\psi,0\rangle$. The state vector is used to provide a conditional probability.  For the \textit{practical} use of quantum theory, the conditional probability
 is the following: given  an apparatus designed to measure an eigenvalue of some operator $A$ at time $t$,  the probability the experimental result is $a$ is $|\langle a|\psi,t\rangle|^{2}$.  

However, suppose one wishes to apply the quantum formalism to (measurement-independent) \textit{reality}.  The minimal and natural  generalization of the conditional probability is  the following:  given time $t$,  the probability nature possesses some eigenvalue $a$ of some operator $A$ is $|\langle a|\psi,t\rangle|^{2}$.  

The problem with this generalization (often called the``measurement problem," but which I prefer to call the``reality problem") is two-fold. 

First, there is nothing in the quantum formalism which chooses one operator over another. One may call this the ``preferred basis problem" since, expressed in terms of the operator's eigenvectors, no basis is ``preferred" over another. That is, with equal justification, one could state: given time $t$, the probability nature possesses some eigenvalue $b$ of some operator $B$ (which does not commute with $A$) is $|\langle b|\psi,t\rangle|^{2}$. 

Suppose, however, that one could add a sensible preferred basis as an extra postulate to quantum theory (something that has not yet been achieved, except in restricted examples).

Then there is yet a second problem, which one may call the ``hopping problem."  The theory gives the joint conditional probability of $a$ at time $t$ and of $a'$ at time $t'$, namely   $|\langle a|\psi,t\rangle|^{2}|\langle a'|\psi,t'\rangle|^{2}$.  However, even if $t'$ is infinitesimally close to $t$, $a'$ can be different from $a$.  That is,  probabilities of states at every instant are all that is specified.   Probabilities of transitions from one state to another as time evolves are not specified.  There is nothing to say that reality cannot do a discontinuous ``hop" from one eigenvalue to another.  

It is because of the preferred basis problem that the founders of quantum theory restricted it to ``measurement situations."   An ill-defined concept\cite{Bell}, nonetheless, the restriction  allows choice of  unarguably real states. 

It is because of the hopping problem that the founders of quantum theory invoked the collapse postulate.  An ill-defined concept\cite{firstpaper}, nonetheless,  invoking collapse ensures that a system remains in a well-defined physical state, at least for a while. 

But, in fundamental (non-measurement) applications, such as attempting to apply quantum theory to the universe, these two problems are there.  One needs to add a preferred basis to quantum theory and one needs a means of overcoming hopping. 

It should be emphasized that we are supposing quantum theory applies to  a single real entity, not to an ensemble of real entities. One may suppose that quantum theory applies only to an ensemble of identical systems\cite{Leslie}. 

In that case, the preferred basis problem remains: what are the states occupied by the ensemble? And, at first sight, it might seem that hopping is no longer a problem since $|\langle a|\psi,t\rangle|^{2}|\langle a'|\psi,t'\rangle|^{2}$ is interpreted as the fraction of the ensemble for which $a$ is the value at $t$ and $a'$ is the value at time $t'$.  However, nothing in the ensemble interpretation precludes hopping for the individual elements of the ensemble such that the overall statistics are unaffected. 

Finally, one may not wish to adopt the ensemble interpretation because it means giving up the description of individual reality which we know exists.  

In discussing the probabilities associated to possible states of an individual, one may relate them to the frequencies of occurrence of the possible states of a hypothetical ensemble,  with the understanding that only one of the elements of the ensemble actually exists.  (Thus, odds are given on a single horserace.) This is not to adopt the ensemble interpretation, which asserts that the ensemble is not hypothetical, that all of its elements exist.

\subsection{Collapse Quantum Theory}\label{sec1B}

The CSL (Continuous Spontaneous Localization) dynamical collapse theory
(\cite{collapseP,collapseGPR}: reviews in \cite{collapsereviews PP,collapsereview G}), of which a simpler formulation is to be utilized here,  provides a resolution of these problems.   

The state vector is meant to correspond to reality.  In order to achieve this, the Schr\"odinger equation is altered by adding an anti-hermitian ``collapse hamiltonian" to the usual hamiltonian. 
The collapse hamiltonian depends upon the (commuting) set of mass-density operators.  Their joint eigenstates constitute the preferred basis, selected in this way for this theory.    

A state vector rapidly evolves  to an approximate eigenstate of mass density when that density is large: such states essentially describe what we see when we look at nature.  When the mass density is small, the dynamics scarcely affects the standard quantum evolution,  as for an isolated particle.  

Although the state vector evolution is toward a mass-density eigenstate, it never gets there.   To get there would be to put particles in eigenstates of position and so infinite energy. This is prevented by the hamiltonian part of the evolution, so that a stasis is reached: for example, the center of mass wave function of an object reaches an equilibrium size, with the hamiltonian evolution tending to spread it, and the collapse hamiltonian evolution tending to narrow it. 

When the spread in an eigenvalue is sufficiently small, one may say that nature possesses that value\cite{collapsereviews PP}. For a measurement situation, where the state vector is initially a superposition of measurement outcome states, the state vector rapidly evolves to essentially one of those states, with the Born rule probability. 

How can solving the Schr\"odinger equation give rise to different outcomes?  The collapse hamiltonian depends upon a fluctuating c-number field $w({\bf x}, t)$,  which is of white noise character (i.e., it can take on any value at any time at any place).    
The state vector $|\psi,t\rangle_{w}$, evolving non-unitarily from the initial state, depends upon $w({\bf x}, t)$: $|\psi,t\rangle_{w}=U'(t, w)|\psi,0\rangle$.  Different $w({\bf x}, t)$'s lead to different results.  It is supposed that nature supplies a particular $w({\bf x}, t)$, but we don't know which one. 
 
Therefore, in addition to the altered Schr\"odinger equation, the theory contains a second equation.  It gives the probability that nature chooses  a particular $w({\bf x},t)$ in an  infinitesimal range.  This probability is $Dw_{w}\langle\psi,t|\psi, t\rangle_{w}$ ($Dw$ is a suitably defined infinitesimal, e.g.,  see Eq. (\ref{10})).  Since the evolution is non-unitary, the squared state vector norm is a dynamical quantity. 
  Thus, large norm state vectors are more likely than small norm state vectors.   Of  course, the dynamics is such that the integral of the probability  over all possible $w$'s up to time $t$ is $\int Dw_{w}\langle\psi,t|\psi,t\rangle_{w}=1$.  
  
  These two equations, the altered Schr\"odinger equation and the probability rule, completely specify the theory.

 \subsubsection{Hopping}\label{sec1B1}
 
Probability theory requires specifying a sample space and probability measure. For standard quantum theory, if a preferred basis $|a\rangle$ has been chosen,  that is the sample space.  The probability measure at time $t$ associated to each $|a\rangle$ is the squared projection $|\langle a|\psi,t\rangle|^{2}$.  As we have explained, this admits hopping. 

In CSL, the  situation is quite different.  The sample space is the set of all $|\psi,t\rangle_{w}$'s.  The probability measure at time $t$ associated to a $|\psi,t\rangle_{w}$ is that of the $w$ which governed the evolution of that  state vector.  

In particular, the squared projection in CSL has no probabilistic interpretation.  When $a$ corresponds to a macroscopic quantity, then $|\langle a|\psi,t\rangle_{w}|^{2}/_{w}\langle \psi,t|\psi,t\rangle_{w}$ is either close to 1 or 0  (the normalization factor is needed since the state $|\psi,t\rangle_{w}$ is not normalized).  In that case, if one wishes, one may use a truncated sample space, grouping  $|\psi,t\rangle_{w}$'s, which are ``close" to $|a\rangle$, as the element $|a\rangle$, and treating the total probability of all the associated $w$'s as the associated probability measure. 

The meaning of probability (the connection with observation) is that one imagines a hypothetical ensemble evolving under all possible $w$'s.  Then, the probability measure associated to $|a\rangle$ is equated to the fraction of those evolutions where $a$ would be observed. 

Since the probability $Dw_{w}\langle\psi,t|\psi, t\rangle_{w}$ is that of a state vector's whole trajectory under the influence of a specific $w(t')$ up to time $t$, not just that of the state vector at the instant $t$, the probability of hopping can be calculated.    Such a calculation shall be given in Section \ref{sec2B2}, where it shall be seen that  continuity is provided by Brownian motion, $B( t)\equiv\int_{0}^{t}dt'w(t')$, which makes hopping negligible. 

For all so-far performed experiments, the predictions of standard quantum theory and CSL theory are indistinguishable.  However, there are so-far unperformed experiments for which they differ, enabling tests to decide between the two\cite{expt}.

\section{Chronogenesis Model}\label{sec2} 

First we present the model of time creation, then we shall discuss its interpretation.

As mentioned in section \ref{sec1}, we start with Eq. (\ref{1}).  

Time shall be discrete. The time operator 
  $T$ has eigenvalues $m\tau$  (said to denote an instant) where $m$ is an integer and $\tau$ is  a fundamental unit of time which, for definiteness, we shall take as the Planck time $\tau$.  We choose  $T\equiv\tau M$ where $M\equiv b^{\dagger}b$. $b^{\dagger}$, $b$  are creation and annihilation operators obeying the standard relations: $[b,b^{\dagger}]=1$ with  $b|0\rangle_{{\cal T}}=0$, and basis vectors  $|m\rangle\equiv b^{\dagger m}|0\rangle_{{\cal T}}/{m!}^{1/2}$.  $V$ depends upon $b^{\dagger}$, and so does not commute with $T$.    

$V$ also depends upon an operator $A$ which shall be related to the size of the universe. 
$A$  shall be taken to commute with $T$,  so  a joint basis $|m\rangle| a\rangle$ exists, where each basis state  $|a\rangle$ corresponds to a different universe size.  For simplicity of discussion, in this section only, we shall suppose that 
the $|a\rangle$ basis is $L$-dimensional.  Therefore, the nothingness state can be written as the direct product of a state of no time and a state of no space, $|0\rangle=|0\rangle_{{\cal T}}|0\rangle_{{\cal S}}$.  

We also assume that the universe has a finite maximum  age, $(K-1)\tau$, with $K$ a given large integer.

Here is our choice for $V$:
\begin{equation}\label{2}
V\equiv\frac{ \Theta(K-M)}{K^{1/2}}\frac{1}{1-\frac{1}{M^{1/2}}e^{-\tau G(M)}e^{-i\tau H(M)}b^{\dagger}}.  
\end{equation}
\noindent $H(M)$ is to be the hamiltonian. $G(M)$ is to be the collapse hamiltonian.  The step function in Eq. (\ref{2}) ensures the maximum age of the universe. The purpose of the $K^{1/2}$ normalization factor will appear during the interpretation discussion. The $M^{1/2}$ factor ensures that the time eigenstates are properly normalized. 

Putting Eq. (\ref{2}) into Eq. (\ref{1}) yields
\begin{equation}\label{3}
|\Psi\rangle=\frac{\Theta(K-M)}{K^{1/2}}\sum_{m=0}^{\infty}\Big[\frac{1}{M^{1/2}}e^{-\tau G(M)}e^{-i\tau H(M)}b^{\dagger}\Big]^{m}|0\rangle.
\end{equation}
\noindent We note that $F(M)b^{\dagger j}=b^{\dagger j}F(M+j)$, where $F$ is an arbitrary function of $M$ (here applied to $G(M)$, $H(M)$ and $M^{1/2}$).  It follows, for the product of $m$ terms, that
\begin{eqnarray}
F(M)b^{\dagger}...F(M)b^{\dagger}  |0\rangle_{{\cal T}}&=&b^{\dagger m}F(M+m)...F(M+1)  |0\rangle_{{\cal T}}\nonumber\\
&=&\sqrt{m!} |m\rangle F(m)...F(1)\nonumber.  
\end{eqnarray}
\noindent Applied to Eq. (\ref{3}), we obtain
\begin{eqnarray}\label{4}
|\Psi\rangle&=&K^{-1/2}\sum_{m=0}^{K-1}e^{-\tau G(m)}e^{-i\tau H(m)}e^{-\tau G(m-1)}e^{-i\tau H(m-1)}\nonumber\\
&&\qquad\qquad\qquad ...e^{-\tau G(1)}e^{-i\tau H(1)} |m\rangle |0\rangle_{{\cal S}}.
\end{eqnarray}

\subsection{Case of No Collapse, $G=0$}\label{sec 2A}

First we consider the case of no collapse,  $G(m)=0$  Then, Eq. (\ref{4}) yields
\begin{equation}\label{5}
K^{1/2}\langle m|\Psi\rangle=e^{-i\tau H(m)}e^{-i\tau H(m-1)} ...e^{-i\tau H(1)}  |0\rangle_{{\cal S}}.  
\end{equation}
This  is clearly the discrete-time version of $|\psi,t\rangle$, the solution to the usual Schr\"odinger equation with time-dependent hamiltonian $H(t/\tau)$: in particular, note the crucial appropriate time-ordering.  

The state vector in Eq. (\ref{5}) is a superposition of the $L$ states $|a\rangle$.

To begin our discussion of interpretation, let each state  corresponding to an instant  (i.e., each term in the sum in Eq. (\ref{4})), be an element of the sample space. Let the probability measure on each state be its squared norm which, according to Eq. (\ref{5}), is 
\begin{equation}\label{6}
\langle\Psi|m\rangle\langle m|\Psi\rangle=\frac{1}{K}.
\end{equation}

How is  this to be interpreted?   In standard quantum theory, when the state vector is a superposition of states, the states are interpreted as alternative possibilities, only one of which is realized.  But that does not hold in this case, since all instants are realized.  

To see what must be done, consider some horse race analogies. 

For an analogy to standard quantum theory, suppose one goes to a racecourse where there are  $L$ horses competing in just one race.    The $L$ horses make up the sample space, and one may assign probabilities, according to the oddsmaker's odds. A probability would  correspond to a measure of one's best subjective judgment\cite{Cox} about the chance of the associated horse winning the race. Or, one may imagine a hypothetical ensemble of races, and then the  probability would correspond to the fraction of total races the corresponding horse wins. 

This is not what we  want. 

For an analogy to our model, suppose that, at the racecourse, there are instead to be $K$ races, numbered $m= 0,...K-1$.  However, each race has only one horse running, chosen according to some statistical rule.   Since the horse that runs is the horse that wins,   it would not make sense to bet on which horse wins.  Still, one could still  bet upon which horse runs in each race.  

The sample space has $KL$ elements, each labeled by a race number  and a horse and an associated probability measure $P(m,n)$.  Considering a  hypothetical ensemble of  race days, $P(m,n)$ would correspond to the fraction of total races represented by the $m$th race run by the $n$th horse. 

The marginal $\sum_{n=1}^{L}P(m,n)=1/K$ is the fraction of total races represented by the $m$th race, regardless of which horse runs. Of course, this not only applies to the hypothetical ensemble of race days, it also applies to the actual race day.  One may imagine people falling asleep before the races begin, waking up while one race is in progress, and using the probability  $1/K$ as guidance to wager on which numbered race they are seeing. 

The marginal $\sum_{m=0}^{K-1}P(m,n)$ is the fraction of total races run by the $n$th horse, regardless of which races it runs in. That too could provide a probability for a wager. 

To summarize,  the probability theory structure of sample space and measure, usually applied to a  circumstance of unsure alternative outcomes of a single event, may also be applied to a circumstance of multiple sure events, each with unsure alternative outcomes.

That is the way we interpret our example.  What corresponds to the hypothetical ensemble of race days is a hypothetical ensemble of universe histories, only one of which is actually realized.  What corresponds to the $K$ races are the $K$ instants.  What corresponds to the $L$ horses are the possible sizes of the universe.

With the operator $A$  representing the volume of space (a precise definition shall be given in the next section),  we shall suppose $H(M)$ does not in fact depend upon $M$, i.e.,   $H$ is time-independent, but $H$ does depend upon $A$ and possibly other operators.

We cannot have  $[A,H]=0$.  Then there is a joint eigenbasis of $A$ and $H$. Since the initial state $|0\rangle_{{\cal S}}$ of no space is then a non-degenerate eigenstate of $A$,  it is an eigenstate of  $H$.  This means that the state vector will not evolve out of the initial state.  

Therefore, since we wish the spatial part of the universe to grow, we must choose $[A,H]\neq 0$.  Then, the spatial state (\ref{5}) at the $m$th instant
 will indeed be a superposition of different $|a\rangle$ states, i.e., of different size volumes of space. 

One expects such a superposition of  ``different universes"  in a quantum theory of universe creation.     The sample space is the joint eigenbasis $|a\rangle|m\rangle$, describing  space at the $m$th instant as having a definite size.  The probability measure on this joint eigenbasis is $|\langle a|\langle m|\Psi\rangle|^{2}$: it is clear from (\ref{5}) that
\[
\sum_{a,m}|\langle a|\langle m|\Psi\rangle|^{2}=\sum_{m=0}^{K-1}\frac{1}{K}=1. 
\]

This  joint probability measure is interpreted, for the hypothetical ensemble of identical  universes, as the fraction of total instants  where the $m$th instant \textit{and} size $a$ jointly occur.  The marginals $\sum_{a}|\langle a|\langle m|\Psi\rangle|^{2}=1/K$  and $\sum_{m}|\langle a|\langle m|\Psi\rangle|^{2}$ correspond respectively to the fraction of total instants 
in the ensemble that the $m$th instant represents and the fraction of total instants  at which the universe size is characterized by $a$. Each universe in the hypothetical ensemble has $K$ instants, so $1/K$ is also the fraction of total instants that the $m$th instant represents for each universe in the ensemble.

In considering this case without collapse, we have done the best we can, but this is not tenable. There is nothing to prevent hopping. Probabilities for $|a\rangle|m\rangle$ and  $|a'\rangle|m+1\rangle$, are given, where the universe sizes $a$ and $a'$ can be completely different.  Something is needed to say that a universe cannot have such a sequence.  

\subsection{Case of Collapse, $H=0$}\label{sec II B}
In Eq. \ref{4}, we shall choose as collapse hamiltonian
\begin{equation}\label{7}
G(m)=\frac{1}{4\lambda}[w(m)-2\lambda A]^{2}.
\end{equation}
\noindent In the remainder of this section, we set $H=0$ so we can discuss the collapse motion  undisturbed by the hamiltonian motion. 

We shall see here that the collapse dynamics tends toward the eigenstates of the operator $A$ in 
Eq.(\ref{7}). We shall choose $A\equiv N\equiv a^{\dagger}a $, where $N$ represents the number of fundamental volume units in the universe which, for definiteness, we may take as 
the Planck volume $\ell^{3}$, i.e., the cube of the Planck length.   
 $a^{\dagger}$, and $a$ are the usual creation and annihilation operators, with $a|0\rangle_{{\cal S}}=0$ and basis vectors  $|n\rangle\equiv a^{\dagger n}|0\rangle_{{\cal S}}/{n!}^{1/2}$.  
 
 The picture is that, due to the combined effects of the hamiltonian and the collapse hamiltonian, space is a ``clump" which grows by adding Planck volume ``lumps."   However, in this section, since $H=0$, there is no creation of space.  Therefore,  instead of the initial state $ |0\rangle_{{\cal S}}$, we shall replace it in Eq.(\ref{4}) by $|\psi, 0\rangle_{{\cal S}}\equiv\sum_{n}\alpha_{n}|n\rangle$.  That is, we consider collapse dynamics acting on an initial state which is a superposition of different universe sizes.
 
In Eq.(\ref{7}),  $\lambda$ characterizes the collapse rate.  $w(m)$ is a random function, defined on the integers $0\leq m\leq K-1$,  which can take on any value at each integer. To make this dependence explicit, we relabel $|\Psi\rangle$ as $|\Psi\rangle_{w}$. 

The sample space is the set of all $\langle m|\Psi\rangle_{w}$'s for every possible $m$ and $w(m)$. The probability measure associated to each is:  
 \begin{equation}\label{8}
P_{m}(w)D_{m}w\equiv_{w}\negthinspace\negthinspace  \langle \Psi| m\rangle\langle m|\Psi\rangle_{w}\prod_{k=1}^{m}\frac{dw(k)}{\sqrt{2\pi\lambda/\tau}}.
\end{equation}
\noindent Given all possible instants over the life of a universe, and all possible $w$'s over the hypothetical ensemble of universes, the probability (\ref{8}) is the fraction represented by the $m$th instant and a $w$ in an infinitesimal range.  

That the total probability is 1 follows from Eqs.(\ref{4},{7},{8}):
\begin{eqnarray}\label{9}
&&\sum_{m=0}^{K-1}\int_{-\infty}^{\infty}D_{m}wP_{m}(w)=\sum_{m=0}^{K-1}\Bigg[\prod_{k=1}^{m}\int_{-\infty}^{\infty}\frac{dw(k)}{\sqrt{2\pi\lambda/\tau}}\Bigg]\cdot\nonumber\\
&&\frac{1}{K}\ _{{\cal S}}\langle\psi, 0|e^{-\frac{\tau}{2\lambda}{[(w(1)-2\lambda A)^{2}+ ...
(w(m)-2\lambda A)^{2}]}}|\psi, 0\rangle_{{\cal S}}\nonumber\\
&=&\sum_{m=0}^{K-1}\frac{1}{K}=1.
\end{eqnarray}

We now consider, for this  simple case of $H=0$, (1) how the collapse mechanism works, (2) how  ``hopping,"  while possible, has a negligible probability of occurring, and (3) how to analyze the ``spread" of an eigenvalue. 

\subsubsection{Collapse Mechanism}\label{sec II B1}

In what follows, we shall employ a \textit{continuum notation}, replacing $\tau$ by $dt$ and replacing the product of exponentials in Eq.(\ref{4}) by a (time-ordered) integral over the exponent.  When performing $dw(t)$ integrations, we shall revert to the use of discrete values of $t$. However, this notation could rather be regarded as an approximation, neglecting terms of order  $\tau$ compared to $t$.  Additionally, we shall relabel $w(m)$ as $w(t)$, and set $\sqrt{K}\langle m|\Psi\rangle_{w}\equiv |\psi,t\rangle_{w}$ so the probability Eq.(\ref{8}), multiplied by $K$, 
\begin{equation}\label{10}
P_{t}(w)D_{t}w\equiv _{w}\negthinspace\negthinspace\langle \psi,t|\psi,t\rangle_{w}\prod_{t'=dt}^{t}
\frac{dw(t')}{\sqrt{2\pi\lambda/dt}},
\end{equation}
\noindent is the conditional probability of $w$ given $t$, whose total, integrated over all $w$, is 1.  

Using  Eqs.(\ref{4}, \ref{7}), the state vector evolution is

 \begin{equation}\label{11}
|\psi,t\rangle_{w}= e^{-\frac{1}{4\lambda}\int_{0}^{t}dt'[w(t')-2\lambda N]^{2}} \sum_{n}\alpha_{n}|n\rangle.  
\end{equation}

The simplest way to see how this evolution collapses the state vector is as follows. Expand the exponent in Eq.(\ref{11}) and write $B(t)\equiv\int_{0}^{t}dt'w(t')$, so 

\begin{eqnarray}\label{12}
|\psi,t\rangle_{w}&=& \sum_{n}\alpha_{n}|n\rangle e^{-\frac{1}{4\lambda}\int_{0}^{t}dt'w^{2}(t')}e^{B(t)n-\lambda tn^{2}}\nonumber\\
&=&C(t)\sum_{n}\alpha_{n}|n\rangle  
 e^{-\frac{1}{4\lambda t}[B(t)-2\lambda tn]^{2}}, 
\end{eqnarray}
\noindent where
\begin{equation}\label{13}
C(t)\equiv e^{-\frac{1}{4\lambda}\int_{0}^{t}dt'w^{2}(t')}
e^{\frac{1}{4\lambda t}B^{2}(t)}.  
\end{equation}
\noindent Using (\ref{12}), the probability (\ref{10}) is 
\begin{eqnarray}\label{14}
P_{t}(w)D_{t}w&=& C^{2}(t)\prod_{t'=dt}^{t}
\frac{dw(t')}{\sqrt{2\pi\lambda/dt}}\cdot\nonumber\\
 &&\qquad\sum_{n}|\alpha_{n}|^{2} e^{-\frac{1}{2\lambda t}[B(t)-2\lambda tn]^{2}}.
 \end{eqnarray}

The  $n$ dependence  in Eqs.(\ref{12},\ref{14}) involves only $B(t)$. We therefore write 
$w(t')=[B(t')-B(t'-dt)]/dt$ with $B(0)=0$, and change the variables of integration from $w$'s to $B$'s. Integrating $C^{2}(t)$ over all $dB(t')$ for $t'=dt, 2dt, ...t-dt$ (not integrating over  $dB(t)$), we  have 
\begin{eqnarray}\label{15}
&&\int Dw_{t}C^{2}(t)=\frac{dB(t)}{\sqrt{2\pi\lambda dt}}e^{\frac{1}{2\lambda t}B^{2}(t)}\prod_{t'=dt}^{t'=t-dt}\int_{-\infty}^{\infty} \frac{dB(t')}{\sqrt{2\pi\lambda dt}}\cdot\nonumber\\
&&e^{-\frac{1}{2\lambda dt}[B(t)-B(t-dt)]^{2}}...e^{-\frac{1}{2\lambda dt}[B(2dt)-B(dt)]^{2}}e^{-\frac{1}{2\lambda dt}[B(dt)-0]^{2}}\nonumber\\
&&\qquad=\frac{dB(t)}{\sqrt{2\pi\lambda t}}e^{\frac{1}{2\lambda t}B^{2}(t)}e^{-\frac{1}{2\lambda t}B^{2}(t)}
=\frac{dB(t)}{\sqrt{2\pi\lambda t}}
\end{eqnarray}
\noindent where the Gaussian integrals have been performed using
\begin{eqnarray}
&&\int_{-\infty}^{\infty}dy\frac{1}{\sqrt{2\pi\alpha}}e^{-\frac{1}{2\alpha}(x-y)^{2}}\frac{1}{\sqrt{2\pi\beta}}e^{-\frac{1}{2\beta}(y-z)^{2}}=\nonumber\\
&&\qquad\qquad\qquad\frac{1}{\sqrt{2\pi(\alpha+\beta)}}
e^{-\frac{1}{2(\alpha+\beta)}(x-z)^{2}}.\nonumber
\end{eqnarray}

\noindent Therefore, by Eq. (\ref{14}), the probability of $B(t)$ is
\begin{equation}\label{16}
P(B(t))dB(t)=\frac{dB(t)}{\sqrt{2\pi\lambda t}} \sum_{n}|\alpha_{n}|^{2} e^{-\frac{1}{2\lambda t}[B(t)-2\lambda tn]^{2}}.
\end{equation}

The collapse behavior is evident in Eq. (\ref{16}).  The sum describes a set of gaussians drifting apart to the right with different constant speeds.  While the width of the gaussians (characterized by the standard deviation $\sqrt{\lambda t}$)  also grows with time, it does not grow as fast as the gaussians separate.  For sufficiently large $\lambda t$, one may say that the only probable values of $B(t)$ lie ``within"  each gaussian, within the range $2\lambda n t\pm R\sqrt{\lambda t}$, where $R$ is some integer $>>1$. 

For such large $\lambda t$, integration of Eq.(\ref{16}) over $B(t)$ within the gaussian whose drift is  $2\lambda tn'$, shows that the associated probability has the Born rule value $|\alpha_{n'}|^{2}$ to high accuracy.  For such $B(t)$, the associated state vector Eq.(\ref{12}), when properly normalized, is the ``collapsed" state vector 
\begin{eqnarray}\label{17}
\frac{|\psi,t\rangle_{w}}{ \sqrt{_{w}\langle \psi,t|\psi,t\rangle_{w}}}&=&
 \frac{C(t)\sum_{n}\alpha_{n}|n\rangle  
 e^{-\frac{1}{4\lambda t}[B(t)-2\lambda tn]^{2}}}{C(t)[\sum_{n}|\alpha_{n}|^{2} e^{-\frac{1}{2\lambda t}[B(t)-2\lambda tn]^{2}}]^{1/2}},\nonumber\\
&\approx&\frac{\alpha_{n'}}{|\alpha_{n'}|}|n'\rangle.  
\end{eqnarray}
\subsubsection{Negligible Hopping}\label{sec2B2}

Unlike the case of standard quantum theory, one can calculate the joint probability $P_{12}$ that the state vector at time $t_{1}$ is (close to) an eigenstate $n_{1}$ of $N$ and, after an interval  $t_{2}=t-t_{1}$,  is (close to) a different  eigenstate $n_{2}$.  As shall be seen, this probability is very small after $t_{1}$ has become large enough that the gaussians have separated (which shall be assumed for the remainder of this discussion). This is  essentially because $B(t)$ is random walk (with drift), which is not likely to go out of one  gaussian into another once they  are well-separated.  

We obtain the probability that $B$ lies in the intervals $B(t_{1})\pm dB(t_{1})$ at time $t_{1}$ and $B(t)\pm dB(t)$ at time $t$ by integrating Eq.(\ref{14}) over all $B$'s except $B(t_{1})$ and $B(t)$, obtaining  
\begin{eqnarray}\label{18}
&&P(B(t),B(t_{1}) )dB(t)dB(t_{1})=\nonumber\\
&&=\frac{dB(t)dB(t_{1})}{2\pi\lambda\sqrt{t_{2} t_{1}}} \sum_{n}|\alpha_{n}|^{2}\cdot\nonumber\\ 
&&\quad e^{-\frac{1}{2\lambda t_{2}}[B(t)-B(t_{1})-2\lambda t_{2}n ]^{2}}
e^{-\frac{1}{2\lambda t_{1}}[B(t_{1})-2\lambda t_{1}n]^{2}}.
\end{eqnarray}
\noindent (Note that we recover Eq.(\ref{16}) when $B(t_{1})$ is integrated over.) 

As a warmup exercise, we show that $B(t)$ is random walk with drift.  That is,  \textit{within} a gaussian, a change of $B$ over a time interval $dt$ is $\sim \sqrt{dt}$.   In 
Eq.(\ref{18}), write $t_{2}=dt$ and set $B(t_{1})=2\lambda n_{1}t_{1}+x$, $B(t)=2\lambda n_{1}t+y$, obtaining for the joint probability of $x$ and $y$ (neglecting much smaller terms), 
\[
P(y,x)dydx\approx \Bigg[\frac{dy}{\sqrt{2\pi\lambda dt} }e^{-\frac{1}{2\lambda dt}[y-x ]^{2}}\Bigg] \frac{|\alpha_{1}|^{2} dx}{\sqrt{2\pi\lambda t_{1}}} e^{-\frac{1}{2\lambda t_{1}}[x ]^{2}}.
 \]
 \noindent The bracketed term is the conditional probability, given x, that y is reached $dt$ later:  this is indeed random walk with standard deviation $\sqrt{\lambda dt}$.
 
To find $P_{12}$, set $B(t_{1})=2\lambda n_{1}t_{1}+x$ and  $B(t)=2\lambda n_{2}t+y$ in Eq. (\ref{18}): 
\begin{eqnarray}\label{19}
&&P(y,x )dydx\approx\frac{dydx}{2\pi\lambda\sqrt{t_{2} t_{1}}}\cdot \nonumber\\
&&\Big[ |\alpha_{1}|^{2} e^{-\frac{1}{2\lambda t_{2}}[y-x+2\lambda t(n_{2}-n_{1}) ]^{2}}
e^{-\frac{1}{2\lambda t_{1}}x^{2}}\nonumber\\
&&+|\alpha_{2}|^{2}
 e^{-\frac{1}{2\lambda t_{2}}[y-x+2\lambda t_{1}(n_{2}-n_{1}) ]^{2}}
e^{-\frac{1}{2\lambda t_{1}}[x-2\lambda t_{1}(n_{2}-n_{1})]^{2}}\Big].\nonumber \\
\end{eqnarray}
\noindent The other $ |\alpha_{n}|^{2}$ terms make much smaller contributions and so have been neglected. 

We note that if  $B(t_{1})$ and $B(t)$ are at the peaks of their respective gaussians (setting $x=y=0$ in Eq.(\ref{19})), we obtain,
 \begin{eqnarray}\label{20}
 &&P(y,x )=\frac{1}{2\pi\lambda\sqrt{t_{2} t_{1}}}\cdot \nonumber\\
&&\quad\Big[ |\alpha_{1}|^{2} e^{-2\lambda \frac{t^{2}}{t_{2}}(n_{2}-n_{1})^{2}}
+|\alpha_{2}|^{2}
e^{-2\lambda \frac{tt_{1}}{t_{2}}(n_{2}-n_{1})^{2}}\Big]\nonumber\\
&&< \frac{1}{2\pi\lambda\sqrt{t_{2} t_{1}}}\Big[ |\alpha_{1}|^{2} e^{-8\lambda t_{1}(n_{2}-n_{1})^{2}}
+|\alpha_{2}|^{2}
e^{-2\lambda t_{1}(n_{2}-n_{1})^{2}}\Big].\nonumber \\
\end{eqnarray}
\noindent The inequality in Eq.(\ref{20}) comes from setting $t_{2}= t_{1}$ in the first exponent and $t_{2}=\infty$ in the second exponent (those are the values for which the magnitudes of the exponents are smallest).  Since the condition for the gaussians to be well-separated at time $t_{1}$ is $\lambda t_{1}|n_{2}-n_{1}|>>R\sqrt{\lambda t_{1}}$, or $\lambda t_{1}[n_{2}-n_{1}]^{2}>>R^{2}$, where $R$ is some suitably large integer, we see that  the probability is quite small for such a hop.  

More generally, we calculate the probability $P_{12}$ that $B( t_{1})$ and $B(t)$ lie anywhere within their respective  gaussians by integrating  Eq.(\ref{19}) over $x$ for the interval $\pm R\sqrt{\lambda t_{1}}$ and over  $y$ for the interval $\pm R\sqrt{\lambda t_{2}}$.  We shall assume, for simplicity, that  $t_{2}>>t_{1}$. Then  we may neglect $x$ in the first exponent in the first term in the bracket, and $x-2\lambda t_{1}(n_{2}-n_{1})$ in the first exponent in the second term in the bracket, since their spreads $\approx R\sqrt{\lambda t_{1}}$ are  much less than the spreads of the remaining terms $\approx R\sqrt{\lambda t_{2}}$.   
Integrating over $x$ in the first term and $y$ in the second term we are left to calculate (using $t_{2}\approx t$):

\begin{eqnarray}\label{21}
&&P_{12}\approx |\alpha_{1}|^{2}\int_{-R\sqrt{\lambda t_{2}}}^{R\sqrt{\lambda t_{2}}}\frac{dy}{\sqrt{2\pi\lambda t_{2}}} e^{-\frac{1}{2\lambda t_{2}}[y+2\lambda t_{2}(n_{2}-n_{1}) ]^{2}}\nonumber\\
&&\qquad+|\alpha_{2}|^{2}
 \int_{-R\sqrt{\lambda t_{1}}}^{R\sqrt{\lambda t_{1}}}\frac{dx}{\sqrt{2\pi\lambda t_{1}}} e^{-\frac{1}{2\lambda t_{1}}[x-2\lambda t_{1}(n_{2}-n_{1}) ]^{2}}\nonumber\\
 &\approx&\sum_{k=1}^{2}\frac{\sinh\Big[2(n_{2}-n_{1})R\sqrt{\lambda t_{k}}\Big]}{\sqrt{\pi\lambda  t_{k}}(n_{2}-n_{1})}
 |\alpha_{k}|^{2}e^{-2\lambda  t_{k}(n_{2}-n_{1})^{2}},
 \end{eqnarray}
\noindent where the integrals have been performed by neglecting $x^{2}$ and $y^{2}$ compared to the much larger remaining terms. We again see, from the last term in Eq.(\ref{21}), that the probability is quite small for such a hop.  
\subsubsection{Spread of $N$}

The state vector is an entangled state, the sum of states  $|m\rangle|\psi,t\rangle_{w}$ which describe an instant and  the physical state of the universe at that instant.  If this is  to make physical sense, for each universe in the hypothetical ensemble,  $|\psi,t\rangle_{w}$ must be close to an eigenstate of the operator $N$ which characterizes the size of the universe.  That is,  the standard deviation of  $N$ for each individual universe, governed by its own particular $w(t)$, should be ``small."  

We use the simplified notation 
$\langle O \rangle_{w}\equiv_{w}\negthinspace\negthinspace\langle\psi,t| O|\psi,t\rangle_{w}$
\noindent and $\overline{F}\equiv\int Dw F(w)$  where $O$ is any operator and $F$ is any function of $w$. Then, the squared standard deviation of $N$ for a particular $w$ can be written as 
\begin{equation}\label{22}
\sigma_{N}^{2}(w)\equiv \frac{\langle\Big[N-\frac{\langle N\rangle_{w}}{\langle 1\rangle_{w}}\Big]^{2}\rangle_{w}}{\langle 1 \rangle_{w}}=\frac{\langle N^{2}\rangle_{w}}{\langle 1 \rangle_{w}}-
\frac{\langle N\rangle_{w}^{2}}{\langle 1 \rangle_{w}^{2}}.
\end{equation}
\noindent (The factor $\langle 1 \rangle_{w}$ is needed to normalize the unnormalized vectors 
$|\psi,t\rangle_{w}$.) Therefore, using Eq.(\ref{8}), the ensemble averaged squared standard deviation of $N$ is given by 
\begin{equation}\label{23}
\overline{\sigma_{N}^{2}}=\int Dw\langle 1 \rangle_{w}\sigma_{N}^{2}(w)=\overline{\langle N^{2}\rangle}-
\overline{\Bigg[\frac{\langle N\rangle^{2}}{\langle 1 \rangle}\Bigg]}.
\end{equation}
\noindent 

We note the sequence of inequalities
\begin{equation}\label{24}
\overline{\langle N^{2}\rangle}\geq\overline{\Bigg[\frac{\langle N\rangle^{2}}{\langle 1 \rangle}\Bigg]}\geq\overline{\langle N\rangle}^{2},
\end{equation}
 \noindent the last following from
 \[
 0\leq  \int  Dw \langle 1 \rangle_{w} \Bigg[\frac{\langle N\rangle_{w}}{\langle 1 \rangle_{w}} - \overline{\langle N\rangle}\Bigg]^{2} =    \overline{\Bigg[\frac{\langle N\rangle^{2}}{\langle 1 \rangle}\Bigg]}-\overline{\langle N\rangle}^{2}.    
 \]
 
 The first term on the right hand side of Eq.(\ref{23}) is quadratic in $|\psi,t\rangle_{w}$ and is generally straightforward to calculate, while the second term is not. However, both can readily be found for the simple case $H=0$ which we have been discussing.  We can find the first term using Eqs.(\ref{12}, \ref{15}), 
 \[
 \overline{\langle f(N)\rangle}=\int Dw_{w}\langle \psi,t| f(N) |\psi,t\rangle_{w}=\sum_{n}|\alpha_{n}|^{2} f(n). 
 \]
 \noindent so 
 \begin{equation}\label{25}
 \quad\overline{\langle N^{2 }\rangle}=\sum_{n}|\alpha_{n}|^{2}n^{2}.
\end{equation}
\noindent The second term is
\begin{eqnarray}\label{26}
&&\overline{\Bigg[\frac{\langle N\rangle^{2}}{\langle 1 \rangle}\Bigg]}=
\int_{-\infty}^{\infty}\frac{dB(t)}{\sqrt{2\pi\lambda t}}\cdot\nonumber\\
&& \frac{\sum_{n,m}|\alpha_{n}|^{2}{n}|\alpha_{m}|^{2}{m} e^{-\frac{1}{2\lambda t}[B(t)-2\lambda tn]^{2}}
e^{-\frac{1}{2\lambda t}[B(t)-2\lambda tm]^{2}}}{\sum_{n}|\alpha_{n}|^{2} e^{-\frac{1}{2\lambda t}[B(t)-2\lambda tn]^{2}}}\nonumber\\
&&\qquad\approx\sum_{n}|\alpha_{n}|^{2}n^{2}=\overline{\langle N^{2 }\rangle}
\end{eqnarray}
\noindent where the limit $\lambda t>>R\sqrt{\lambda t}$ has been taken in the last step. As $t\rightarrow\infty$, the result (\ref{26}) becomes exact.  

Thus, by Eq.(\ref{23}), $\overline{\sigma_{N}^{2}}\rightarrow 0$ in this limit: for each $w$, the spread of $N$ asymptotically vanishes,  indicating complete collapse to a particular size for each universe.  

When we consider $H\neq 0$,  such simple behavior is not obtained.  However, as shall be shown, in the model of the next section, the collapse provides a sufficiently narrow spread of size for each universe. 

\section{Cosmogenesis Model}\label{sec3} 

To describe the creation of space, we choose $H=\epsilon a^{\dagger}a+g(a+a^{\dagger})$. 
$n\epsilon$ is the energy of the volume of space comprised of $n$ fundamental volumes, and $g$ characterizes the growth rate of space. With the collapse hamiltonian (\ref{7}),  the state vector evolution is
\begin{equation}\label{27}
|\psi,t\rangle_{w}= {\cal T}e^{-\int_{0}^{t}dt'[i\epsilon a^{\dagger}a+ig(a+a^{\dagger})+\frac{1}{4\lambda}(w(t')-2\lambda a^{\dagger}a)^{2}]}|0\rangle_{{\cal S}},
\end{equation}
\noindent where ${\cal T}$ is the time-ordering operator. 

We have introduced this model earlier\cite{PCosmo}, but the calculation of $\overline{\sigma_{N}^{2}}$ and the methods used here have not been presented before.  

This model has some interesting aspects.  

\textit{Without} collapse,  $\langle\psi,t |N|\psi,t\rangle$ doesn't grow. It  simply oscillates from 0 and returns, since the hamiltonian is  that of a displaced harmonic oscillator.  

\textit{With} collapse, as shall be seen, asymptotically, $\overline{\langle N\rangle}\sim t$. Thus, in this model,  collapse is responsible for the growth of the universe! Moreover, the linear behavior allows $\overline{\langle N\rangle}$ to be used as a physical clock, correlated to the eigenvalues of $T$.  
 
 The growth of $\overline{\langle N\rangle}$ occurs because the oscillation periodically creates eigenstates with eigenvalues higher than 0, the collapse then tends to enhance one or another of these eigenstates, the oscillation proceeds to higher eigenvalues from there, etc.  
 
[Of course, one can make models where the growth of the universe does not require collapse  For example, suppose $\epsilon=0$ in all subsequent equations. The Hamiltonian is then just $\sim$ the position operator,  a.k.a.  the momentum translation generator. Then, the universe grows, with or without collapse.  Without collapse (i.e., $\lambda=0$),  $\overline{\langle N\rangle}=(gt)^{2}$. With collapse, as shall be seen, the growth is slowed down to asymptotically  linear in time.]  
  
 We shall now exhibit the asymptotic behavior, and then show that  the spread is negligible for sufficiently large $t$, i.e., $\sqrt{\overline{\sigma_{N}^{2}}}/\overline{\langle N\rangle}_{\overrightarrow {t\rightarrow\infty}}0$, which is crucial for a consistent physical picture. 
  
 \subsection{ Statevector }
 Using the Fourier transform relation
 \begin{eqnarray}\label{28}
&&\int_{-\infty}^{\infty}\frac{d\eta(t')}{\sqrt{\pi/\lambda dt }}e^{-dt\lambda \eta^{2}(t')}
e^{idt\eta(t')[w(t')-2\lambda N]}=\nonumber\\
&&\qquad\qquad e^{-dt\frac{1}{4\lambda}(w(t')-2\lambda N)^{2}},
\end{eqnarray}
\noindent Eq. (\ref{27}) may be written
\begin{eqnarray}\label{29}
|\psi,t\rangle_{w}&=& \int D\eta e^{-\lambda\int_{0}^{t}dt' \eta^{2}(t')}e^{i\int_{0}^{t}dt\eta(t')w(t')}\cdot\nonumber\\
&&\quad{\cal T}e^{-i\int_{0}^{t}dt'[(\epsilon +2\lambda\eta(t'))N+g(a+a^{\dagger})]}|0\rangle_{{\cal S}}.
\end{eqnarray}
\noindent Now that the quadratic $N$ dependence has been made linear, the time-ordered expression  can easily be evaluated, resulting in  
\begin{equation}\label{30}
|\psi,t\rangle_{w}= \int D\eta e^{-\lambda\int_{0}^{t}dt \eta^{2}(t')}e^{i\int_{0}^{t}dt'\eta(t')w(t')}
 e^{\gamma(t)+\alpha(t)a^{\dagger}}|0\rangle_{{\cal S}}
\end{equation}
\noindent where
\begin{equation}\label{31}
\frac{d}{dt}\alpha(t)=-ig-i[\epsilon+2\lambda\eta(t)]\alpha     \hbox{ and } \frac{d}{dt}\gamma(t)=-ig\alpha
\end{equation}
\noindent so
\begin{equation}\label{32}
\alpha(t)=-ig\int_{0}^{t}dt'e^{-i\int_{0}^{t}dt''[\epsilon+2\lambda\eta(t'')]\theta(t''-t')},
\end{equation}
\noindent where $\theta$ is the step function. Therefore, 
\begin{eqnarray}\label{33}
&&|\langle n|\psi, t\rangle_{w}|^{2}=\int D\eta'\int D\eta  e^{-\lambda\int_{0}^{t}dt'[ \eta'^{2}(t')+\eta^{2}(t')]}\nonumber\\
&&
e^{-i\int_{0}^{t}dt[\eta'(t')-\eta(t')]w(t')}e^{\gamma'^{*}(t)+\gamma(t)}\frac{[\alpha'^{*}(t)\alpha(t)]^{n}}{n!},
\end{eqnarray}
\noindent where the prime (no prime) on $\alpha$, $\gamma$ means that the associated quantity depends upon $\eta'$ ($\eta$). It follows  from Eq. (\ref{31}): 
\begin{equation}\label{34}
\frac{d}{dt}[\gamma'^{*}(t)+\gamma(t)+\alpha'^{*}(t)\alpha(t)]=2i\lambda[\eta'(t)-\eta(t)]\alpha'^{*}(t)\alpha(t),
\end{equation}
\noindent so Eq. (\ref{33}) can be written as
\begin{eqnarray}\label{35}
&&|\langle n|\psi, t\rangle_{w}|^{2}=\int D\eta'\int D\eta  e^{-\lambda\int_{0}^{t}dt'[ \eta'^{2}+\eta^{2}]}\nonumber\\
&&\negthickspace\negthickspace e^{-i\int_{0}^{t}dt'[\eta'-\eta][w-2\lambda\alpha'^{*}\alpha]}
\frac{[\alpha'^{*}(t)\alpha(t)]^{n}}{n!}e^{-\alpha'^{*}(t)\alpha(t)}\nonumber\\
\end{eqnarray}

\subsection{Calculation of $\overline{\langle N\rangle}$ and $\overline{\langle N^{2}\rangle}$}
 Since the integral over $Dw$ in Eqs.(\ref{35}) results in $\prod_{t'=0}^{t}\delta[\eta'(t')-\eta(t')]$, we have
 \begin{eqnarray}\label{36}
&&\overline{\langle N\rangle}=\sum_{n=0}^{\infty}n\int Dw|\langle n|\psi, t\rangle_{w}|^{2}\nonumber\\
&&=\int D\eta  e^{-2\lambda\int_{0}^{t}dt'\eta^{2}(t')} \alpha^{*}(t)\alpha(t)
\end{eqnarray}
\noindent  Using $ \alpha$ given by Eq.(\ref{32}), and performing the integral over $\eta$, we obtain

 \begin{eqnarray}\label{37}
 &&\overline{\langle N\rangle}=\nonumber\\
&&g^{2}\int_{0}^{t}dt_{1}\int_{0}^{t}dt_{2}
e^{-i\epsilon (t_{2}-t_{1})}e^{-\frac{\lambda}{2}
\int_{0}^{t}dt''[\theta(t''-t_{1})-\theta(t''-t_{2})]^{2}}\nonumber\\
&&=\frac{g^{2}}{\epsilon^{2}+(\lambda/2)^{2}}[\lambda t+2e^{-\lambda t/2}\cos(\epsilon t+2\phi)-
2\cos2\phi],\nonumber\\
\end{eqnarray}
\noindent where $\tan\phi=2\epsilon/\lambda$. 

We see, as mentioned,  that $\overline{\langle N\rangle}$ increases linearly with $t$ for large $t$. When $\lambda$ gets very large, 
the linear growth rate gets very small: this is an example of ``watched pot" or ``Zeno's paradox" behavior\cite{Sud}.  Of course, when there is no collapse, $\lambda=0$, this  
behavior disappears, to be replaced by just oscillation. (When $\lambda$ and $\epsilon$ both vanish, so the hamiltonian is just the position operator, then $\overline{\langle N\rangle}=(gt)^{2}$.)

 Similarly, 
  \begin{eqnarray}\label{38}
&& \overline{\langle N^{2}\rangle}=\sum_{n=0}^{\infty}n^{2}\int Dw|\langle n|\psi, t\rangle_{w}|^{2}\nonumber\\
&&=\int D\eta  e^{-2\lambda\int_{0}^{t}dt'\eta^{2}(t')} [(\alpha^{*}(t)\alpha(t))^{2}+ \alpha^{*}(t)\alpha(t)].
\end{eqnarray}
\noindent    The second term in  the bracket in Eq.(\ref{38}) has been evaluated in Eq.(\ref{37}).  The first term is rather complicated, but we shall only need its  large $t$ behavior, which  is
 \begin{eqnarray}\label{39}
&&\int D\eta  e^{-2\lambda\int_{0}^{t}dt'\eta^{2}(t')} (\alpha^{*}(t)\alpha(t))^{2}=\nonumber\\
&&g^{4}\int_{0}^{t}dt_{1}dt_{2}dt_{3}dt_{4}e^{-i\epsilon (t_{4}+t_{3}-t_{2}-t_{1})}\cdot\nonumber\\
&&e^{-\frac{\lambda}{2}\int_{0}^{t}dt''[\theta(t''-t_{1})+\theta(t''-t_{2})-\theta(t''-t_{3})-\theta(t''-t_{4})]^{2}}\nonumber\\
&&\qquad=2\Big[\frac{g^{2}\lambda t}{\epsilon^{2}+(\lambda/2)^{2}}\Big]^{2}+o(g^{4}t). 
\end{eqnarray}
\noindent Thus, asymptotically,  $\overline{\langle N^{2}\rangle}=2\overline{\langle N\rangle}^{2}+o(t)$.  

Although this result means that
\[ 
 \overline{\langle N^{2}\rangle}-\overline{\langle N\rangle}^{2}= \overline{\langle N\rangle}^{2}+o(t), 
\]
\noindent this does \textit{not} mean that the 
spread of $N$ is comparable to its mean value, since $\overline{\sigma_{N}^{2}}$ is not equal to this expression. 
This expression is the  squared deviation of the ensemble average value of $N$, while $\overline{\sigma_{N}^{2}}$ is the ensemble average of the squared deviation of $N$, which was given in 
Eq.(\ref{23}). In Appendix A is it shown that 
\begin{equation}\label{40}
\overline{\Bigg[\frac{\langle N\rangle^{2}}{\langle 1 \rangle}\Bigg]}=2\overline{\langle N\rangle}^{2}+o(t)
\end{equation}  
\noindent so 
\begin{equation}\label{41}
\overline{\sigma_{N}^{2}}=\overline{\langle N^{2}\rangle}-
\overline{\Bigg[\frac{\langle N\rangle^{2}}{\langle 1 \rangle}\Bigg]}=o(t)
\end{equation}  
and therefore $\sqrt{\overline{\sigma_{N}^{2}}}/\overline{\langle N\rangle}\sim t^{-1/2}\thinspace_{\overrightarrow {t\rightarrow\infty}}0$.

For example, since  $\overline{\sigma_{N}^{2}}\sim t$ and $\overline{\langle N\rangle}\sim t$, then $\overline{\sigma_{N}^{2}}={ \cal C}\overline{\langle N\rangle}$, where 
$\cal{C}$ is some function of  $\epsilon$, $g$, $\lambda$. 
 Denote the volume of the universe by $R^{3}$ and the uncertainty in volume by $(\Delta R)^{3}$. Then, we may estimate  $\Delta R$ as follows. 

With $\ell$ the Planck length, the mean number of Planck volumes is $\overline{\langle N\rangle}\approx (R/\ell)^{3}$.  The uncertainty in the number of Planck volumes is $\approx (\Delta R/\ell)^{3}$, so $\overline{\sigma_{N}^{2}}\approx (\Delta R/\ell)^{6}$. Equating the two expressions for  $\overline{\sigma_{N}^{2}}$, we obtain $\Delta R\approx{\cal C}^{1/6}\sqrt{R\ell}$. For a universe of ``length" $R=c\times 10$ billion years, 
we obtain $\Delta R\approx {\cal C}^{1/6}4\times10^{-3}$cm.

\section{Concluding Remarks}

We have presented a simple model where the state vector $|\Psi\rangle$, describing the universe throughout its history, is given by an operator acting on an initial state of no time and no space, and we have looked at interpretational issues which arise.  We have shown that the sample space/probability measure appropriate to this model is satisfactory, and that it can be sensibly interpreted provided dynamical collapse is involved.  

The elements of the sample space are the states $|m\rangle|\psi, t\rangle_{w}$, where $t=m\tau$. Each state $|m\rangle$ describes an instant. Each state  $|\psi, t\rangle_{w}$ (evolving in a particular universe under the influence of a particular $w(t)$) describes the universe size with a narrow spread.  
The probability assigned to each element is $K^{-1}|_{w} \langle \psi,t|\psi, t\rangle_{w}|^{2}Dw$ and, of course, the total probability satisfies $\sum_{m=0}^{K-1}\int Dw |_{w} \langle \psi,t|\psi, t\rangle_{w}|^{2}=1$. 

Further development of this model would involve adding a hamiltonian and collapse hamiltonian to describe the contents of the universe such as fields.  These hamiltonians would depend upon the volume of space operator $N\ell^{3}$.  This would result in dynamics of space altered from that presented here, e.g., one might introduce gravitational, inflationary, accelerative behavior.    

In this model, there appears to be an asymmetry between time and space. For each element of the sample space,  there is uncertainty in universe size but no uncertainty in time.  
However, the asymmetry disappears if one compares the size of the universe to the age of the universe as measured by a clock.   

A clock may readily be introduced employing the dynamics already discussed. Clock time shall be represented by an operator $T'=M'\tau$ which has a collapse dynamics. We write $M'\equiv a'^{\dagger}a'$,  with creation and annihilation operators $a'^{\dagger}$ and $a'$. The initial state of the clock satisfies  
$a'|0\rangle_{{\cal T}'} =0$, with clock instants represented by the states $|m'\rangle$. We shall call $T'$  ``prime-time"  and  $|m'\rangle$``prime-instants"  to ensure no confusion with $T$ and  $|m\rangle$.

 We introduce a classical random field $w'(t)$ to govern the evolution of prime-time.  The treatment shall be exactly parallel to 
that of space, with hamiltonian and collapse hamiltonian in Eq.(\ref{3}) now given by 
\begin{eqnarray}\label{42}
H&=&\epsilon a^{\dagger}a+g(a^{\dagger}+a)+\epsilon' a'^{\dagger}a'+g'(a'^{\dagger}+a'),\nonumber\\
G&=&\frac{1}{4\lambda}[w(T)-2\lambda a^{\dagger}a]^{2}+\frac{1}{4\lambda'}[w'(T')-2\lambda' a'^{\dagger}a']^{2},\nonumber\\
\end{eqnarray}
\noindent and initial state 
\[
|0\rangle=|0\rangle_{\cal T}|0\rangle_{\cal T'}|0\rangle_{\cal S}.  
\] 
The state vector Eq.(\ref{4}) becomes (using the continuum notation mentioned in the paragraph before Eq.(\ref{10})):
\begin{equation}\label{43}
|\Psi\rangle=K^{-1/2}\sum_{m=0}^{K-1}|m\rangle|\psi, t\rangle_{w}|\psi', t\rangle_{w'},
\end{equation}
\noindent  where $|\psi, t\rangle_{w}$ is given by Eq.(\ref{27}).  $|\psi', t\rangle_{w'}$ is given by Eq.(\ref{27}) with all variables primed and 
$|0\rangle_{\cal S}$ replaced by $|0\rangle_{\cal T'}$.

The sample space elements are  
$|m\rangle|\psi, t\rangle_{w}|\psi', t\rangle_{w'}$  ($t=m\tau$).   
Associated to each element is the probability $K^{-1}| \langle_{w} \psi,t|\psi, t\rangle_{w}|^{2}Dw| \langle_{w'} \psi',t|\psi', t\rangle_{w'}|^{2}Dw'$. Of course, the total probability satisfies $\sum_{m=0}^{K-1}\int Dw |_{w} \langle \psi,t|\psi, t\rangle_{w}|^{2}\int Dw' |_{w'} \langle \psi',t|\psi', t\rangle_{w'}|^{2}=1$. 

The behavior of prime-time is precisely parallel to the behavior of space we have previously discussed. According to the primed version of Eq.(\ref{37}), 
\[
 \overline{\langle M'\rangle}\thinspace_{\overrightarrow {t\rightarrow\infty}}
\frac{g'^{2}\lambda' t}{\epsilon'^{2}+(\lambda'/2)^{2}},    
\]
\noindent i.e., the mean prime-instant, and therefore the mean prime-time $\overline{\langle M'\rangle}\tau$,   grows asymptotically linearly with $t$.
And, according to the primed version of Eq.(\ref{41}), the spread $\overline{\sigma_{M'}^{2}}\sim t\sim \overline{\langle M'\rangle}$. 

We can see the clock's  behavior in the 
position basis starting from the state vector expressed
in the prime-instant basis (Eq.(\ref{A5}) with replacements $n \rightarrow m'$, $\overline{\langle N\rangle}\rightarrow \overline{\langle M'\rangle}$, $w\rightarrow w'$), which is approximately a superposition of equal amplitude states for $\overline{\langle M'\rangle}-\sqrt{\overline{\langle M'\rangle}}\lesssim m'\lesssim \overline{\langle M'\rangle}-\sqrt{\overline{\langle M'\rangle}}$. 
The clock's ``pointer packet" oscillates back and forth with frequency $\epsilon'$ with increasing amplitude $\sim \sqrt{ \overline{\langle M'\rangle}}$.

The uncertainty in the clock's measurement of universe age may be estimated as follows.  We write  $\overline{\sigma_{M'}^{2}}={\cal C'}\overline{\langle M'\rangle}$, where 
$\cal{C}'$ is some function of  $\epsilon'$, $g'$, $\lambda'$. Denote the  age of the universe by ${\cal T}'$ and its uncertainty  by $\Delta {\cal T}'$. 
With $\tau$ the Planck time, the present number of Planck times is $\overline{\langle M'\rangle}\approx{\cal T}'/\tau$.  The uncertainty in the number of Planck times is $\approx \Delta {\cal T}'/\tau$, so $\overline{\sigma_{M'}^{2}}\approx (\Delta {\cal T}'/\tau)^{2}$. Equating the two expressions for  $\overline{\sigma_{M'}^{2}}$, we obtain 
$\Delta {\cal T}'\approx{\cal C}^{1/2}\sqrt{{\cal T}'\/\tau}$. For a universe of ``age" $ {\cal T}'=10$ billion years, 
 $\Delta R\approx {\cal C}'^{1/2}10^{-13}$sec.   

Each sample space element $|m\rangle|\psi, t\rangle_{w}|\psi', t\rangle_{w'}$ corresponds to an instant of specific universe size and age (within the narrow spreads).  The ensemble describes all possible universes for all possible instants. 

Knowing the present universe size and age, and wishing  to describe its history, one truncates the sample space, obtaining the relative probability measure consistent with that knowledge.  The resulting subspace  provides a description of a universe with well-defined evolving size and age.   

\appendix

\section{Calculation of $\overline{\langle N\rangle^{2}/\langle 1 \rangle}$}

We begin with the expression for $|\langle n|\psi, t\rangle_{w}|^{2}$given in Eq.(\ref{35}).  The Poisson distribution appears there.   For large $n$, the Poisson distribution $[x^{n}/n!]e^{-x}$ can be replaced by its  gaussian approximation $(2\pi x)^{-1/2}\exp-[(n-x)^{2}/2x]$, which is good for large $x,n$.   Since it is the large $t$ limit which is of interest, and large $n$'s are excited for large $t$, this is a good approximation. We make this replacement in Eq.(\ref{35}), obtaining 
\begin{eqnarray}\label{A1}
&&|\langle n|\psi, t\rangle_{w}|^{2}
\approx \int D\eta'\int D\eta  e^{-\lambda\int_{0}^{t}dt'[ \eta'^{2}+\eta^{2}]}\nonumber\\
&&\negthickspace\negthickspace e^{-i\int_{0}^{t}dt'[\eta'-\eta][w-2\lambda\alpha'^{*}\alpha]}
\frac{1}{\sqrt{2\pi\alpha'^{*}\alpha}}e^{-\frac{[n-\alpha'^{*}\alpha]^{2}}{2\alpha'^{*}\alpha}}.
\end{eqnarray}

We note that this approximation does not change the calculated values of $\overline{\langle N\rangle}$ and $\overline{\langle N^{2}\rangle}$, since  
\begin{eqnarray}\label{A2}
 \overline{\langle N\rangle}&\approx&\int_{0}^{\infty}ndn\int Dw|\langle n|\psi, t\rangle_{w}|^{2}\nonumber\\
 &\approx&\int_{-\infty}^{\infty}ndn  \int D\eta e^{-2\lambda\int_{0}^{t}dt'\eta^{2}(t')} \frac{1}{\sqrt{2\pi\alpha'^{*}\alpha}}e^{-\frac{[n-\alpha'^{*}\alpha]^{2}}{2\alpha'^{*}\alpha}} \nonumber\\
 &=&\int D\eta  e^{-2\lambda\int_{0}^{t}dt'\eta^{2}(t')} \alpha^{*}(t)\alpha(t)
\end{eqnarray}
\noindent which is Eq.(\ref{36}),  and similarly 
\begin{eqnarray}\label{A3}
 \overline{\langle N^{2}\rangle}&\approx&\int_{0}^{\infty}n^{2}dn\int Dw|\langle n|\psi, t\rangle_{w}|^{2}\nonumber\\
 &\approx&\int_{-\infty}^{\infty}n^{2}dn  \int D\eta e^{-2\lambda\int_{0}^{t}dt'\eta^{2}(t')} \frac{1}{\sqrt{2\pi\alpha'^{*}\alpha}}e^{-\frac{[n-\alpha'^{*}\alpha]^{2}}{2\alpha'^{*}\alpha}}
  \nonumber\\
 &=&\int D\eta  e^{-2\lambda\int_{0}^{t}dt'\eta^{2}(t')} [(\alpha^{*}(t)\alpha(t))^{2}+ \alpha^{*}(t)\alpha(t)]
\end{eqnarray} 
\noindent which is Eq.(\ref{38}).

 	In Eq.(\ref{A1}), because the width of the gaussian $\sqrt{ \alpha'^{*}(t)\alpha(t)}$ is small compared 
 compared to the mean value $\alpha'^{*}(t)\alpha(t)$, we make the approximation  $w-2\lambda\alpha'^{*}\alpha\approx w-2\lambda n$. 
 With this approximation, and  a change of variables to $\xi(t)\equiv \eta (t)-\eta'(t)$, $\mu (t)\equiv \eta (t)+\eta'(t)$,  Eq.(\ref{A1}) becomes
\begin{eqnarray}\label{A4}
&&|\langle n|\psi, t\rangle_{w}|^{2}\approx \int D\xi e^{-\frac{\lambda}{2}\int_{0}^{t}dt'\xi^{2}} e^{-i\int_{0}^{t}dt'\xi [w-2\lambda n]}\cdot\nonumber\\
&&\qquad\int D\mu e^{-\frac{\lambda}{2}\int_{0}^{t}dt'\mu^{2}}\frac{1}{\sqrt{2\pi\alpha'^{*}\alpha}}e^{-\frac{[n-\alpha'^{*}\alpha]^{2}}{2\alpha'^{*}\alpha}}.
\end{eqnarray}

This approximation also does not change the results of the above calculations of $ \overline{\langle N\rangle}$ and $\overline{\langle N^{2}\rangle}$.  
 If the integral over $Dw$ is performed in Eq.(\ref{A4}), the result is $\prod_{t'=0}^{t}\delta[\xi(t')]$. Then, the $D\xi$ integral can  be performed, with the result 
 \[
\int Dw |\langle n|\psi, t\rangle_{w}|^{2}= \int D\mu e^{-\frac{\lambda}{2}\int_{0}^{t}dt'\mu^{2}(t')} \frac{1}{\sqrt{2\pi\alpha'^{*}\alpha}}e^{-\frac{[n-\alpha'^{*}\alpha]^{2}}{2\alpha'^{*}\alpha}}
 \]
 \noindent Setting $\mu=2\eta$, this is identical to the expression for  $\int Dw |\langle n|\psi, t\rangle_{w}|^{2}$ in the middle integral of Eqs.(A2, A3), which gives the correct expressions for $\overline{\langle N\rangle}$ and $\overline{\langle N^{2}\rangle}$.  
 
 Therefore, as our last  approximation, we replace the gaussian in the second integral in Eq.(\ref{A4}) by a gaussian in $n$ which gives these same correct expressions, obtaining
\begin{eqnarray}\label{A5}
&&|\langle n|\psi, t\rangle_{w}|^{2}\approx \int D\xi e^{-\frac{\lambda}{2}\int_{0}^{t}dt'\xi^{2}} e^{-i\int_{0}^{t}dt'\xi [w-2\lambda n]}\cdot\nonumber\\
&&\qquad\qquad\qquad\qquad\qquad\frac{1}{\sqrt{2\pi\sigma^{2}}}
e^{-\frac{\big[n-\overline{\langle N\rangle}\big]^{2}}{2\sigma^{2}}}\nonumber\\
&&=e^{-\frac{1}{2\lambda}\int_{0}^{t}dt'[w(t')-2\lambda n]^{2}}\frac{1}{\sqrt{2\pi\sigma^{2}}}
e^{-\frac{\big[n-\overline{\langle N\rangle}\big]^{2}}{2\sigma^{2}}}
\end{eqnarray}
 \noindent where $\sigma^{2}\equiv \overline{\langle N^{2}\rangle}-\overline{\langle N\rangle}^{2}$.  
 
Now, just as in Section \ref{sec II B1}, we can change variables to $B(t')$'s from $w(t')$'s, and integrate over all $B(t')$'s except  $B(t)$, obtaining the equivalent of Eq. (\ref{A5}), 
 \begin{equation}\label{A6}
|\langle n|\psi, t\rangle_{B}|^{2}\approx \frac{1}{2\pi\sqrt{\lambda t\sigma^{2}}}e^{-\frac{1}{2\lambda t}[B(t)-2\lambda t n]^{2}}e^{-\frac{\big[n-\overline{\langle N\rangle}\big]^{2}}{2\sigma^{2}}}.  
\end{equation}

The rest of the calculation is straightforward.  We immediately find the quantities 
\begin{eqnarray}\label{A7}
\langle 1 \rangle_{B}&=&\int_{-\infty}^{\infty}dn |\langle n|\psi, t\rangle_{B}|^{2}\nonumber\\
&=&\frac{1}{\sqrt{2\pi\lambda t[1+4\lambda t\sigma^{2}]}}
e^{-\frac{\big[B-2\lambda t\overline{\langle N\rangle}\big]^{2}}{\lambda t[1+4\lambda t\sigma^{2}]}}\nonumber\\
&\approx&\frac{1}{\sqrt{2\pi(2\lambda t\sigma)^{2}]}}
e^{-\frac{\big[B-2\lambda t\overline{\langle N\rangle}\big]^{2}}{2[(2\lambda t\sigma)^{2}}}
\end{eqnarray}
\begin{eqnarray}\label{A8}
\langle N\rangle_{B}&=&\int_{-\infty}^{\infty}dn n |\langle n|\psi, t\rangle_{B}|^{2}=\langle 1 \rangle_{B}
\Bigg[\frac{2\sigma^{2}B+\overline{\langle N\rangle}}{1+4\lambda t\sigma^{2}}\Bigg]\nonumber\\
&\approx& \langle 1 \rangle_{B}
\frac{1}{2\lambda t}
 B(t),
\end{eqnarray}
\noindent where 1 has been neglected compared to $4\lambda t \sigma^{2}$. Also,  $\overline{\langle N\rangle}/2\sigma^{2}$, whose contribution vanishes as $t\rightarrow \infty$, has been neglected compared to $B$, whose contribution is infinite in that limit.

Finally, using Eqs. (\ref{A7}, \ref{A8}), we calculate
\begin{eqnarray}\label{A9}
\frac{\overline{\langle N\rangle^{2}}}{\langle 1 \rangle}&=&\int_{-\infty}^{\infty}dB
\Bigg[ \frac{B(t)}{2\lambda t)}\Bigg]^{2}
\frac{1}{\sqrt{2\pi(2\lambda t\sigma)^{2}]}}
e^{-\frac{\big[B-2\lambda t\overline{\langle N\rangle}\big]^{2}}{2[(2\lambda t\sigma)^{2}}}\nonumber\\
&=&\sigma^{2}+\overline{\langle N\rangle}^{2}= \overline{\langle N^{2}\rangle}.
\end{eqnarray}

Although our result is $\overline{\sigma_{N}^{2}}=\overline{\langle N^{2}\rangle}- \overline{\langle N\rangle^{2}/\langle 1 \rangle}     =0$, in view of the approximations made  without assessing the errors involved, it is more conservative to conclude that $\overline{\sigma_{N}^{2}}\sim t$ at worst, which still results in $\overline{\sigma_{N}^{2}}/\overline{\langle N\rangle^{2}}\rightarrow0$ as $t\rightarrow\infty$.

\acknowledgments{ I would like to thank Carl Rubino for help with Greek.}

\end{document}